\documentclass[%
 aip,
 amsmath,amssymb,
reprint
]{revtex4-1}
\usepackage[utf8]{inputenc}
\usepackage[T1]{fontenc}
\usepackage{mathptmx}
\usepackage{etoolbox}
\usepackage{dcolumn}
\usepackage{bm}

\usepackage{subfigure}
\usepackage{graphicx}
\usepackage{upgreek}
\usepackage[section]{placeins}
\usepackage{color}

\makeatletter
\def\@email#1#2{%
 \endgroup
 \patchcmd{\titleblock@produce}
  {\frontmatter@RRAPformat}
  {\frontmatter@RRAPformat{\produce@RRAP{*#1\href{mailto:#2}{#2}}}\frontmatter@RRAPformat}
  {}{}
}%
\makeatother

\begin{document}

\preprint{AIP/123-QED}

\title{A Multiscale Boltzmann Equation for Complex Systems of Neutral Gases across All Flow Regimes}%

\author{Sha Liu}%
\homepage{shaliu@nwpu.edu.cn}
\affiliation{National Key Laboratory of Aircraft Configuration Design, Northwestern Polytechnical University, Xi'an, Shaanxi, China}
\affiliation{Institute of Extreme Mechanics, Northwestern Polytechnical University, Xi'an, Shaanxi, China}
\author{Junzhe Cao}%
\homepage{caojunzhe@mail.nwpu.edu.cn}
\affiliation{National Key Laboratory of Aircraft Configuration Design, Northwestern Polytechnical University, Xi'an, Shaanxi, China}
\author{Sirui Yang}%
\homepage{ysr1997@mail.nwpu.edu.cn}
\affiliation{National Key Laboratory of Aircraft Configuration Design, Northwestern Polytechnical University, Xi'an, Shaanxi, China}
\author{Chengwen Zhong}%
\homepage{Corresponding author: zhongcw@nwpu.edu.cn}
\affiliation{National Key Laboratory of Aircraft Configuration Design, Northwestern Polytechnical University, Xi'an, Shaanxi, China}
\affiliation{Institute of Extreme Mechanics, Northwestern Polytechnical University, Xi'an, Shaanxi, China}

\date{\today}%

\begin{abstract}
A Multiscale Boltzmann Equation (MBE) is found from the gas-kinetic theory as the master equation for complex physical systems of neutral gases across all flow regimes, which span from the continuum limit to the free-molecular limit, covering a vast range of applications such as hypersonic flows over near-space vehicles and delicate flows around micro-electromechanical systems. A key feature of this MBE is the introduction of the observation scale into the master equation, which distinguishes the MBE from single-scale master (or governing) equations where fixed observation scales are implied in their basic assumptions. The fundamental properties of MBE, such as the asymptotic preserving property, are proved theoretically, and a concise numerical scheme is developed for MBE to demonstrate its validity on benchmark multiscale problems.
\end{abstract}

\maketitle
Over the past decade, advancements in complex science and multiscale physics have enabled the identification and investigation of numerous multiscale physical problems in fields such as nuclear physics\cite{nuclear}, plasma physics\cite{plasma1,plasma2}, atmospheric science\cite{atmosphere}, and aerodynamics\cite{turbulance}. The scales of these problems often extend beyond the reach of established theories which address single scale physics only. This leads to knowledge gaps in understanding multiscale physical mechanisms and an absence of definitive multiscale master equations.

In the complex system of neutral gases across all flow regimes from the macroscopic continuum limit to the microscopic free-molecular limit, the challenge of undetermined physical mechanisms is addressed through either the rough averaging of micro-scale information for macro-scale governing equations\cite{r13} or the coupling of microscopic and macroscopic mechanisms using specific physical principles\cite{trmc,apdsmc,emms} and model equations\cite{ugks,dugks,gkua,idvm,ugkwp}. Additionally, some approaches attempt to address this issue through a mathematical synthetic iteration\cite{gsis} of different single scale equations. However, as in many other physical fields about multiscale complex systems, there is a lack of a definitive master equation upon which a concrete theory and numerical methods can be established, providing a stable and rational foundation for these theoretical and numerical treatments. Therefore, we aim to find a multiscale master equation for the complex system of neutral gases in this work.

For neutral gases, continuum flows (near the continuum limit) are described by aerodynamics and the Navier-Stokes (NS) equations, while rarefied flows (near free-molecular limit) are governed by gas kinetic theory and the Boltzmann Equation (BE). Therefore, a multiscale equation should not only bridge the gap between these two limits, but also recover both limits with a concrete asymptotic preserving property\cite{ap}.

As is well known, the BE serves as the master equation of gas-kinetic theory\cite{chapman}, which is a single scale equation whose spatial and temporal scales are the mean free path and the mean collision time of molecules, respectively. Its differential-(nonlinear fivefold) integral mathematical form is as follows:
\begin{equation}\label{equ1}
\frac{{\partial f}}{{\partial t}} + {\bm{\upxi}}\cdot\frac{{\partial f}}{{\partial {\bf{x}}}} = \int_{R^3}\int^{4\pi}_{0}\left({ f^{'}_1f^{'}-f_1f }\right)\xi_r\sigma d\Omega d{\bm{\upxi}}_{1},
\end{equation}
where $f\left({\bf{x}},{\bm{\upxi}},t\right)$ is the molecular distribution, ${\bf{x}}$ is the location, $t$ is the time, ${\bm{\upxi}}$ is the molecular velocity (${\bf{x}}$ and ${\bm{\upxi}}$ are omitted for simplicity in the following pages). $\xi_r=\left|{\bm{\upxi}}-{\bm{\upxi}}_{1}\right|$, $\sigma$ and $\Omega$ are the relative velocity, differential cross section, and solid angle, respectively. The distribution with prime denotes the after-collision one, and the distribution with subscript ``1'' is for velocity ${\bm{\upxi}}_{1}$. The left-hand side of Eq.~\ref{equ1} describes the free-transport of molecules, and the right-hand side is the collision term, briefly noted by $B\left( {f,f} \right)$.

To examine the BE from a multiscale perspective, scale-independent principles are necessary, such as the second law of thermodynamics. Therefore, a multiscale temporal integral solution is introduced, which describes the trend of a random distribution (system) towards its equilibrium state with the maximum entropy, and this process can be written as follows:
\begin{equation}\label{equ2}
f\left( {t + \Delta t} \right) = {e^{ - \frac{{\Delta t}}{\tau }}}f\left( t \right) + \left( {1 - {e^{ - \frac{{\Delta t}}{\tau }}}} \right)g\left( t \right),
\end{equation}
where $\Delta t$ is the observation time, $\tau=\mu/p$ is the relaxation time towards equilibrium, here $\mu$ is the viscosity coefficient, $g$ is the local equilibrium distribution function in the following form:
\begin{equation}\label{equ3}
g = n{\left( {\frac{m}{{2\pi kT}}} \right)^{\frac{{3}}{2}}}\exp \left( { - \frac{{ m\left({\bm{\upxi}}-{\bf{U}}\right)\cdot\left({\bm{\upxi}}-{\bf{U}}\right) }}{{2kT}}} \right),
\end{equation}
where $m$ is the molecular mass, $k$ is the Boltzmann constant. $n$, ${\bf{U}}$, $T$ are macroscopic number density, velocity and temperature. This temporal integral solution can be obtained from the Bhatnagar-Gross-Krook (BGK) model of BE, whose collision term also directly describes the trend of a non-equilibrium distribution towards the equilibrium one.

Since $f\left(t+\Delta t\right)$ from the multiscale temporary integral solution (Eq.~\ref{equ2}) must follow the original BE as an ordinary distribution function, by substituting it into the original BE (Eq.~\ref{equ1}), the master equation turns into:
\begin{equation}\label{equ4}
\begin{aligned}
&e^{ - \Delta t/\tau }\frac{Df}{Dt} + \left( {1 - {e^{ - \Delta t/\tau }}} \right)\frac{Dg}{Dt}\\
& = e^{ - \Delta t/\tau }\left( {1 - {e^{ - \Delta t/\tau }}} \right)\left[ {B\left(g,f\right) + B\left(f,g\right)} \right] \\
& + e^{ - 2\Delta t/\tau }B\left(f,f\right) +  \left( {1 - {e^{ - \Delta t/\tau }}} \right)^2B\left(g,g\right).
\end{aligned}
\end{equation}
Given that the collision term for equilibrium state $g$ vanishes:
\begin{equation}\label{equ5}
\frac{Dg}{Dt} = B\left( {g,g} \right) \equiv 0,
\end{equation}
after simplifications, the resultant equation becomes:
\begin{equation}\label{equ6}
\frac{Df}{Dt} = e^{ - \frac{\Delta t}{\tau }}B\left( {f,f} \right) + \left( {1 - {e^{ - \frac{{\Delta t}}{\tau }}}} \right)\left\{ {B\left( {f,g} \right) + B(g,f)} \right\},
\end{equation}
where $\left\{ {B\left( {f,g} \right) + B(g,f)} \right\}$ is just the collision term of the Linearized Boltzmann Equation (LBE) \cite{cercignani}. By denoting the linearized Boltzmann collision term by $L(f)$, the resultant master equation can be finally written as:
\begin{equation}\label{equ7}
\frac{Df}{Dt} = e^{ - \frac{{\Delta t}}{\tau }}B\left( {f,f} \right) + \left( {1 - {e^{ - \frac{{\Delta t}}{\tau }}}} \right)L\left( f \right).
\end{equation}
The assumption of LBE is that the distribution function is not far from equilibrium, and $L\left(f\right)$ describes the molecular collisions when $\Delta t\gg\tau$ in/near the continuum flow regime. In this work, Eq.~\ref{equ7} is called the Multiscale Boltzmann Equation (MBE), which is obtained through the above process that investigates the microscopic BE from a multiscale perspective. Examining this equation, some useful property can be identified:

\begin{itemize}
\item[$\bm{(1)}$] The observation time $\Delta t$ is introduced into the master equation (Eq.~\ref{equ7}). $\Delta t/\tau$ in the exponential is physically the ratio of observation scale to the molecular transportation scale, where $\Delta t$ spans from microscopic to macroscopic.
\item[$\bm{(2)}$] The multiscale collision term of MBE is a convex combination (weighted average) of the BE one and the LBE one, and it covers the whole flow regime between the continuum limit and free-molecular limit by given different observation scale $\Delta t$ according to research need.
\end{itemize}

It is evident that the MBE has the correct Chapman-Enskog (CE) expansion (asymptotic preserving), since both BE and LBE have this property and MBE is their linear combination. For the same reason, the MBE also satisfies the basic H theorem (the second law of thermodynamics in gas-kinetic theory) and the conservation of collision invariants.

Two types of interpretations of MBE collision term can be made. One interpretation is that molecules follow a weighted collision term during the whole observation time $\Delta t$. This indirectly corroborates some numerical methods for BE in large time step\cite{trmc,apdsmc}. The other interpretation is a two-step process that molecules follow the BE collision term during $\delta t_1=e^{-\Delta t/\tau}\Delta t<\tau$, and follow LBE collision term during $\delta t_2=\left(1-e^{-\Delta t/\tau}\right)\Delta t$. Given the first (order) term of $L\left(f\right)$ is a BGK collision term\cite{cercignani}, $L\left(f\right)$ can be approximated by the relaxation collision term $R(f)=\left(g-f\right)/\tau$ during $\delta t_2$, and the Quantified Model Competition (QMC) mechanism\cite{suwp} for large temporal scale can be used, where a molecule is randomly classified as a free transport one or a colliding one, with the probabilities $e^{ - \frac{{\delta t_2}}{\tau }}$ and $\left( 1 - {e^{ - \frac{{\delta t_2}}{\tau}}} \right)$, respectively, as follows:
\begin{equation}\label{equ8}
f\left( \Delta t \right) = e^{ - \frac{{\delta t_2}}{\tau }}f_{BE} + \left( 1 - {e^{ - \frac{{\delta t_2}}{\tau }}} \right)f_{CE},
\end{equation}
where $f_{BE}$ is the distribution function after $\delta t_1$ and $f_{CE}$ is the second order CE distribution, $f_{CE}=Dg/Dt+{\bm{\upxi}}\cdot\partial g/\partial \bf{x}$. Eq.~\ref{equ8} is obtained by spatial and temporal Taylor expansions of the temporal integral solution of an inhomogeneous BGK equation during $\delta t_2$\cite{suwp}. Eq.~\ref{equ8} can also be viewed as a formal solution of MBE describing that, after an evolution during $\Delta t$, $e^{-\delta t_2/\tau}$ portion of molecules experience a BE process, while $1-e^{-\delta t_2/\tau}$ portion of molecules follow the second order CE distribution corresponding to the macroscopic NS equation.

\begin{figure}[htb]
\centering
\includegraphics[width=0.45\textwidth]{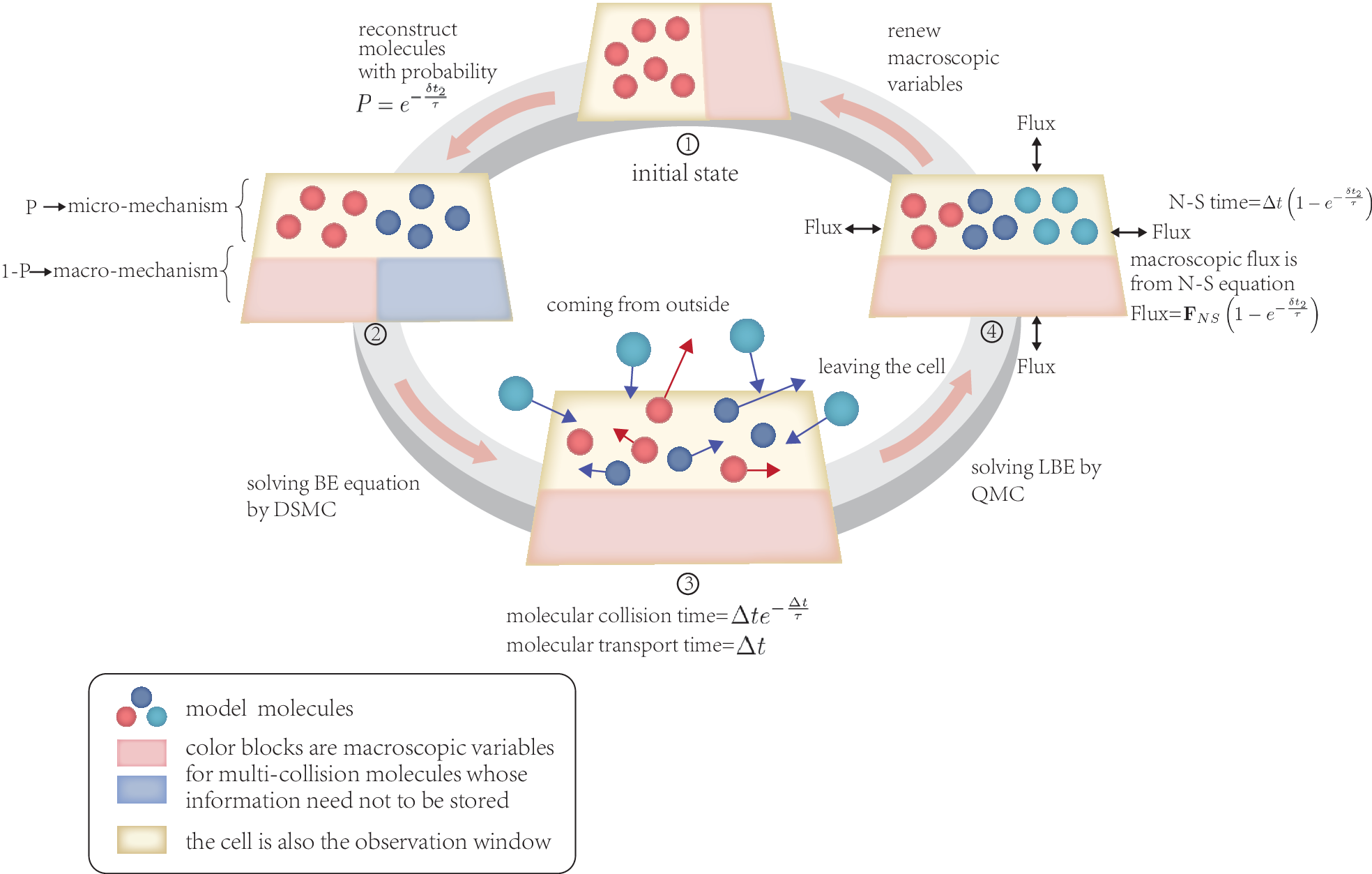}
\caption{\label{Fig:picture} The two-step process for MBE solver.}
\end{figure}

Given the two-step process for the MBE, a concise numerical algorithm for MBE can be formulated as follows, as illustrated in Fig.~\ref{Fig:picture}. The flow field is discretized into discrete cells (volumes). Model molecules are employed to describe the non-equilibrium distribution function $f_{BE}$, each representing a large amount of real molecules\cite{bird}. Meanwhile, the near-equilibrium part of distribution function $f_{CE}$ is entirely determined by macroscopic variables\cite{chapman,suwp}.

\begin{description}
\item[Step 1] At the beginning of a new time step, there are free transport molecules and macroscopic variables representing the colliding molecules from the previous time step. They will be re-categorized in this new time step: The previous free transport molecules are reclassified as candidate free transport and candidate colliding molecules during the large temporal process in $\delta t_2$, with different probabilities (weights) $e^{-\delta t_2/\tau}$ and $1-e^{-\delta t_2/\tau}$. Since the second order CE distribution corresponds to the macroscopic NS equation and is fully determined by the macroscopic variables, the transport of all colliding molecule in cell can be represented by the macroscopic NS equation. Their individual information can be erased and merged into the macroscopic variables. On the other hand, some molecules will emerge from the local Maxwellian distribution determined by the previous macroscopic variables, with a probability $e^{-\delta t_2/\tau}$.

\item[Step 2] For numerically solving the BE equation during $\delta t_1$, the DSMC method\cite{bird} is adopted. This classic rarefied solver involves split and successive free transport and collision processes, and the No-Time Counter collision algorithm is chosen in this work, whose details can be found in Ref.\cite{bird}. In the MBE, this process is followed by an extra free-transport process until the end of time step (during $\delta t_2$). Then, the evolution of microscopic information is complete.

\item[Step 3] The NS equation is solved using a standard Computational Fluid Dynamics (CFD) method\cite{hirsch,leveque} with a modified viscosity coefficient to accommodate multiscale calculation~\cite{suwp}. The overall macroscopic flux of colliding molecules is just the NS numerical flux multiplying by their scale weight $1-e^{-\delta t_2/\tau}$. (The treatment of the flux at the boundary is the same). In this work, the Kinetic Inviscid Flux~\cite{kif} which is convenient for multiscale frameworks is selected.

\item[Step 4] Finally, total macroscopic variables in cell can be updated as follows:

\begin{equation}
\begin{aligned}
&{\bf{W}}^{n + 1} = {\bf{W}}^n - {\bf{W}}_{DSMC}^n + {\bf{W}}_{DSMC}^{n + 1} \\
+& \left({1-e^{-\frac{\delta t_2}{\tau}}}\right)\sum\limits_{cf = 1}^{N_{\rm{face}}} {{\bf{F}}_{NS}} \cdot {{\bf{S}}_{cf}}\Delta t,
\end{aligned}
\end{equation}
\end{description}
where ${\bf{W}}$ is the macroscopic conservation field variables cumulated in cell (volume) $\Omega$ as ${\bf{W}}=\left(\rho\Omega, \rho {\bf{U}}\Omega, \rho U^2\Omega+3\rho RT\Omega\right)^T$, the superscripts $n$ and $n+1$ indicate the iteration time steps. ${\bf{S}}_{cf}$ is the directed area pointing out of a cell. $N_{\rm{face}}$ is the number of cell interfaces. It can be found that a MBE solver only unifies the DSMC and NS solvers by the scale-dependent weights, and both algorithms are almost unchanged.

An important detail should be mentioned: Since numerical meshes often vary significantly in different locations of a flow field, a local and physical time is used to determine the observation scale for each cell, which is $\Delta t_{phy}=L_{cell}/u_{cell}$, where $L_{cell}$ can be the cubic root of a cell volume and $u_{cell}$ can be $|U|+\sqrt{{\gamma}RT}$, where $R=k/m$ is the specific gas constant.  Therefore, $L_{cell}$ and $\Delta t_{phy}$ are the observation length and time scales for this discrete cell, which fits the essence of mesh resolution. $\Delta t_{phy}$ is only used for calculating the scale weights in cells, while the numerical process proceeds in a numerical time step $\Delta t$ which is the smallest $\Delta t_{phy}$ in the entire flow field.

\begin{figure}[htb]
\centering
\subfigure{
\includegraphics[width=0.22\textwidth]{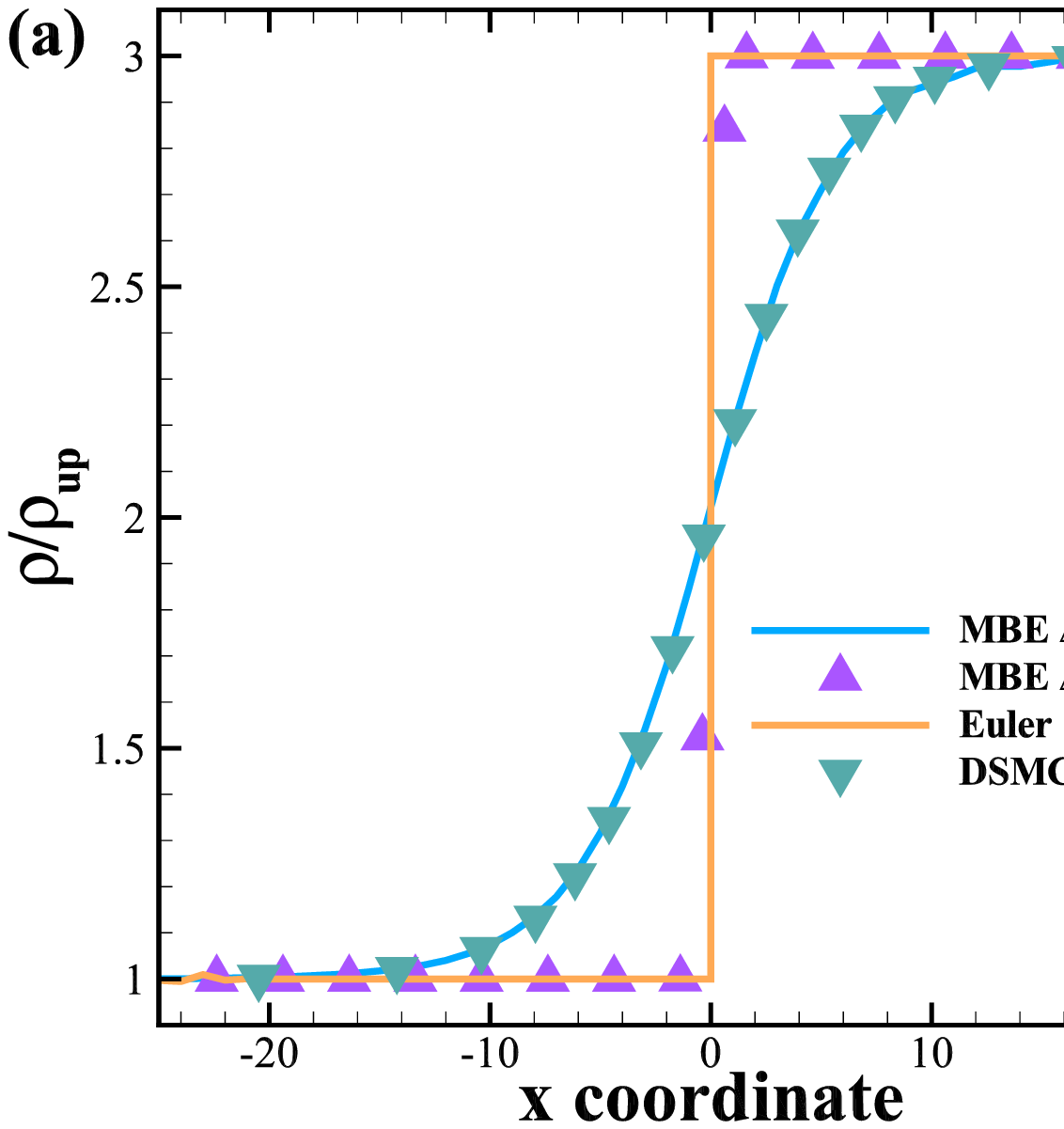}
}\hspace{0.001\textwidth}%
\subfigure{
\includegraphics[width=0.22\textwidth]{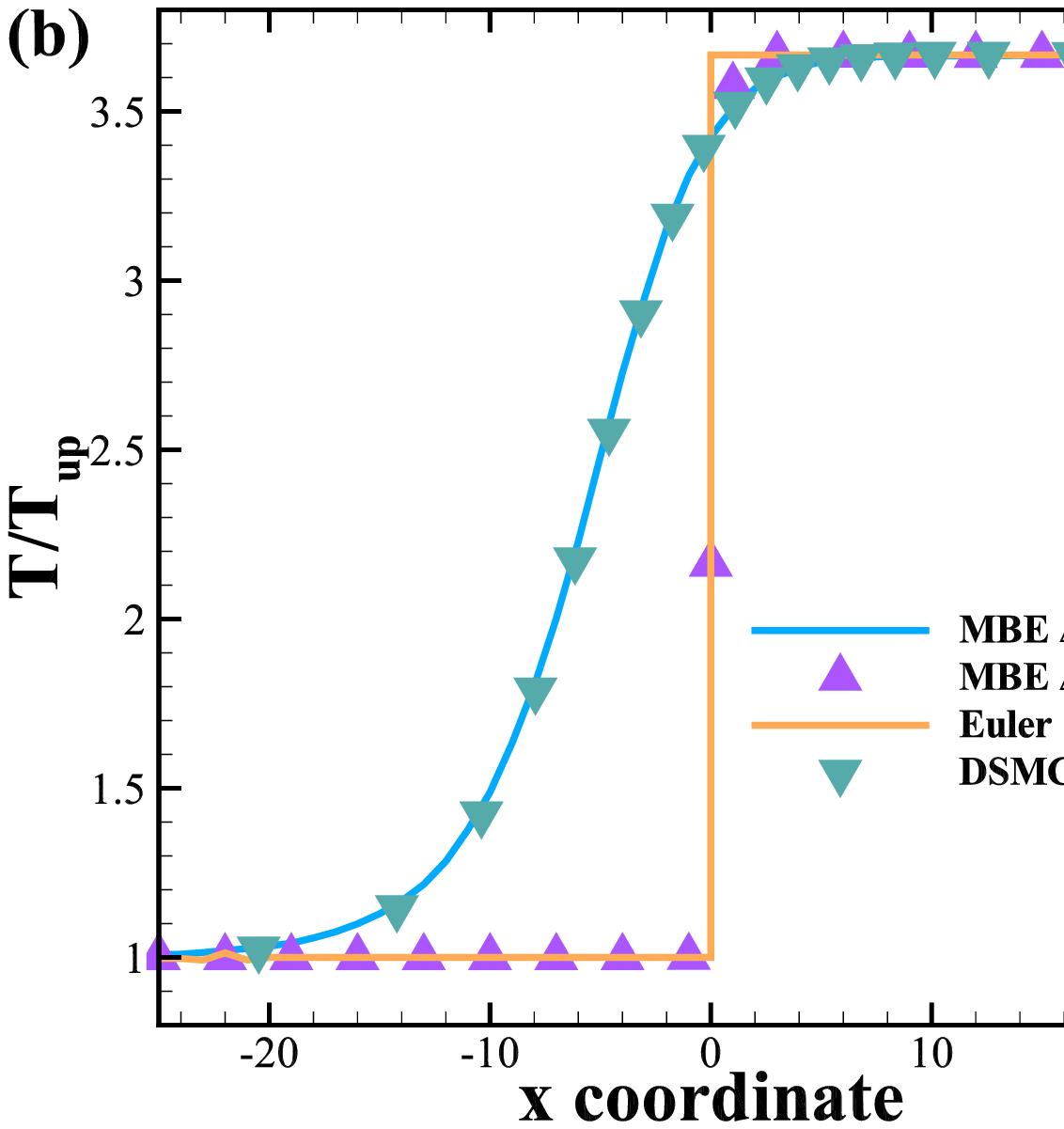}
}
\caption{\label{Fig:fig1} Density and temperature profiles in the internal of a $Ma=3$ shock wave.}
\end{figure}

To test the validity and accuracy of the MBE and its numerical solver, a normal shock wave is calculated from both a macroscopic perspective and a microscopic perspective that zooms into the internal of the shock wave. The upstream Mach number ($Ma=\frac{U_{up}}{\sqrt{{\gamma}R{T_{up}}}}$) is set to be $3$. The argon gas modeled by variable hard sphere potential with heat index $\omega=0.5$ is chosen as the working gas. In Fig.\ref{Fig:fig1}, the MBE solution matches well with the NS results in macro-scale ($\Delta x=10^{5}{{\lambda}}$), and matches well with the DSMC solutions\cite{ohwada} in micro-scale ($\Delta x={{\lambda}}/4$). Here ${{\lambda}}$ is the molecular mean free path in the upstream.

\begin{figure}[htb]
\centering
\subfigure{
\includegraphics[width=0.22\textwidth]{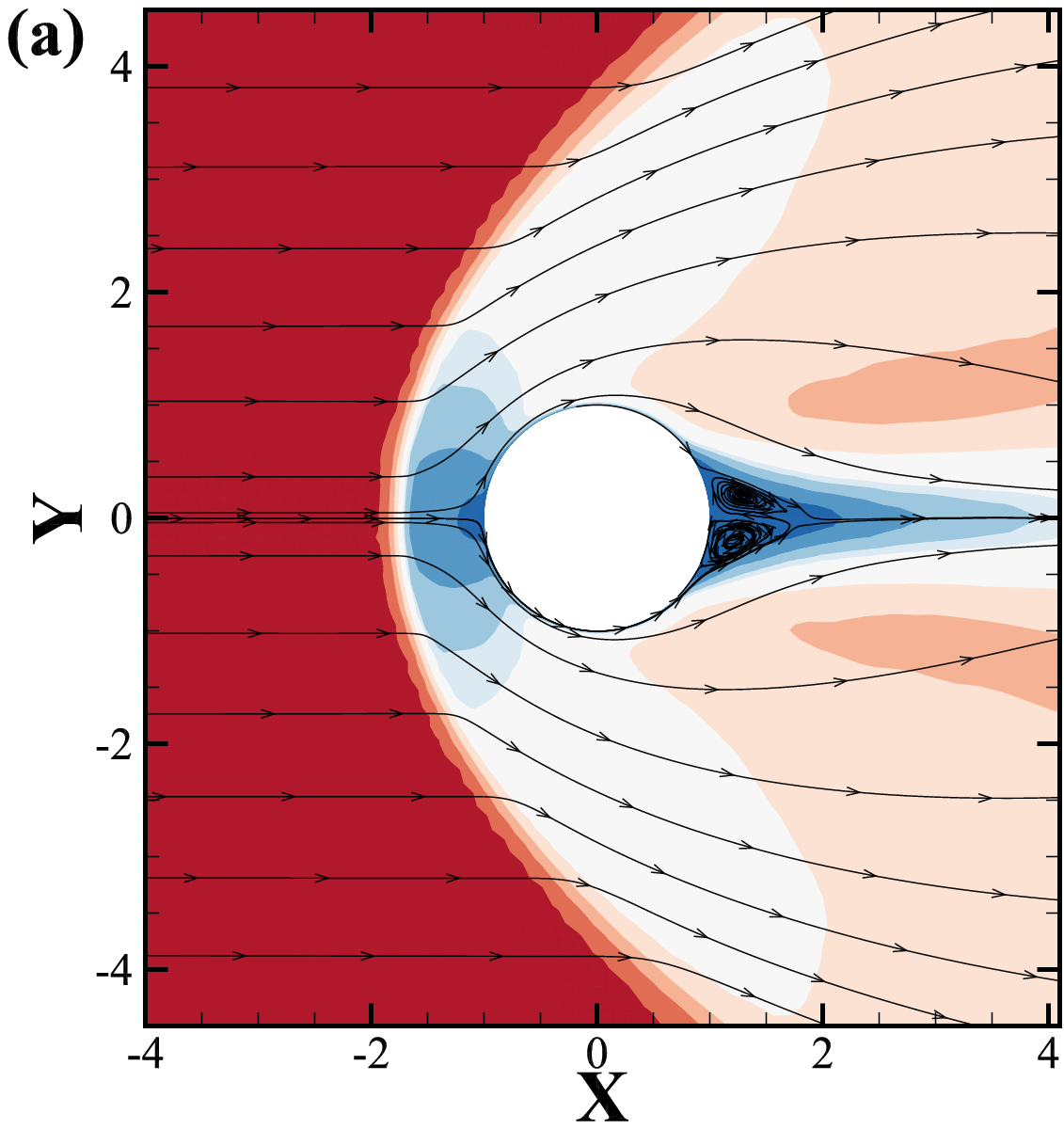}
}\hspace{0.001\textwidth}%
\subfigure{
\includegraphics[width=0.22\textwidth]{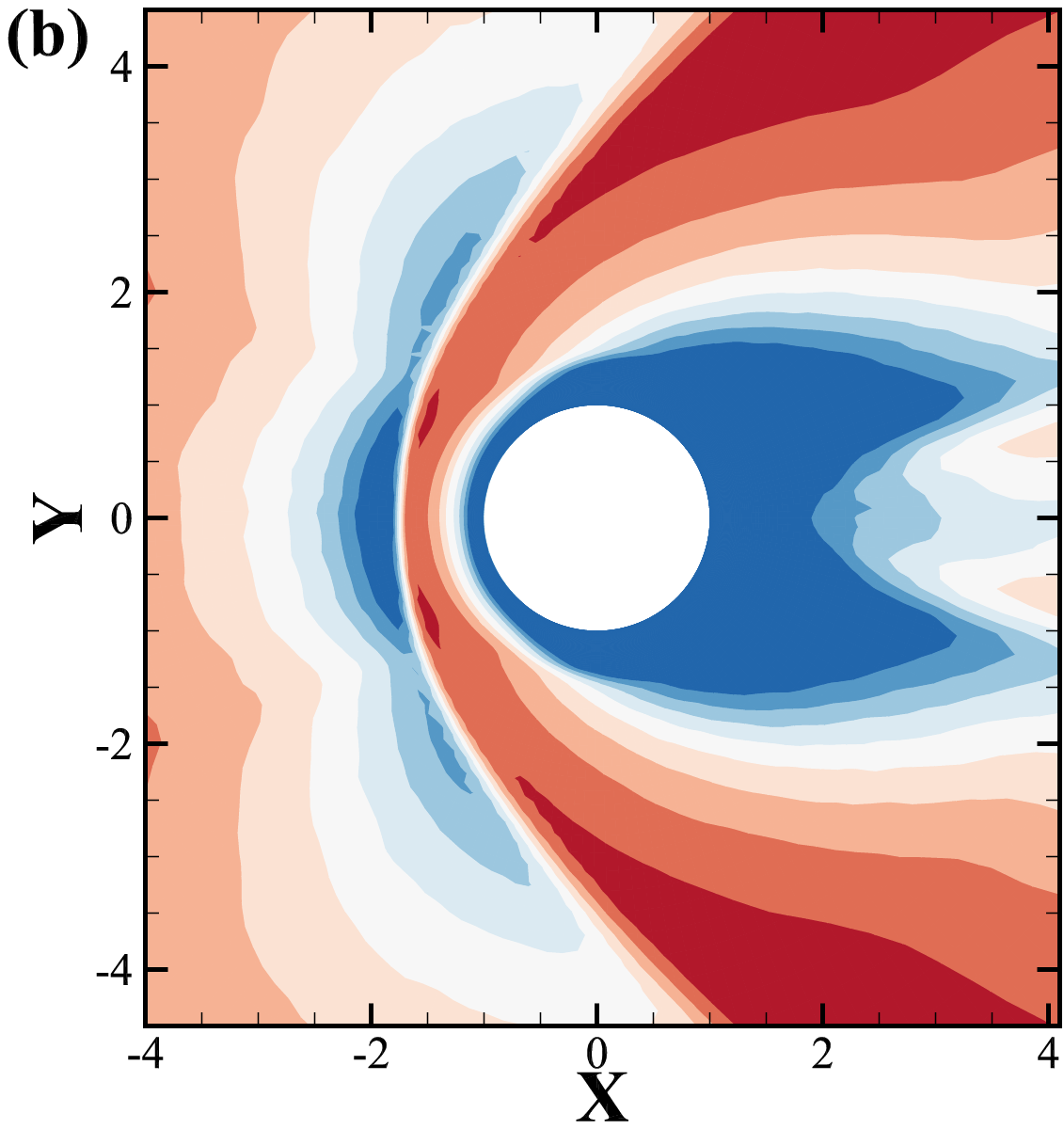}
}\\
\subfigure{
\includegraphics[width=0.22\textwidth]{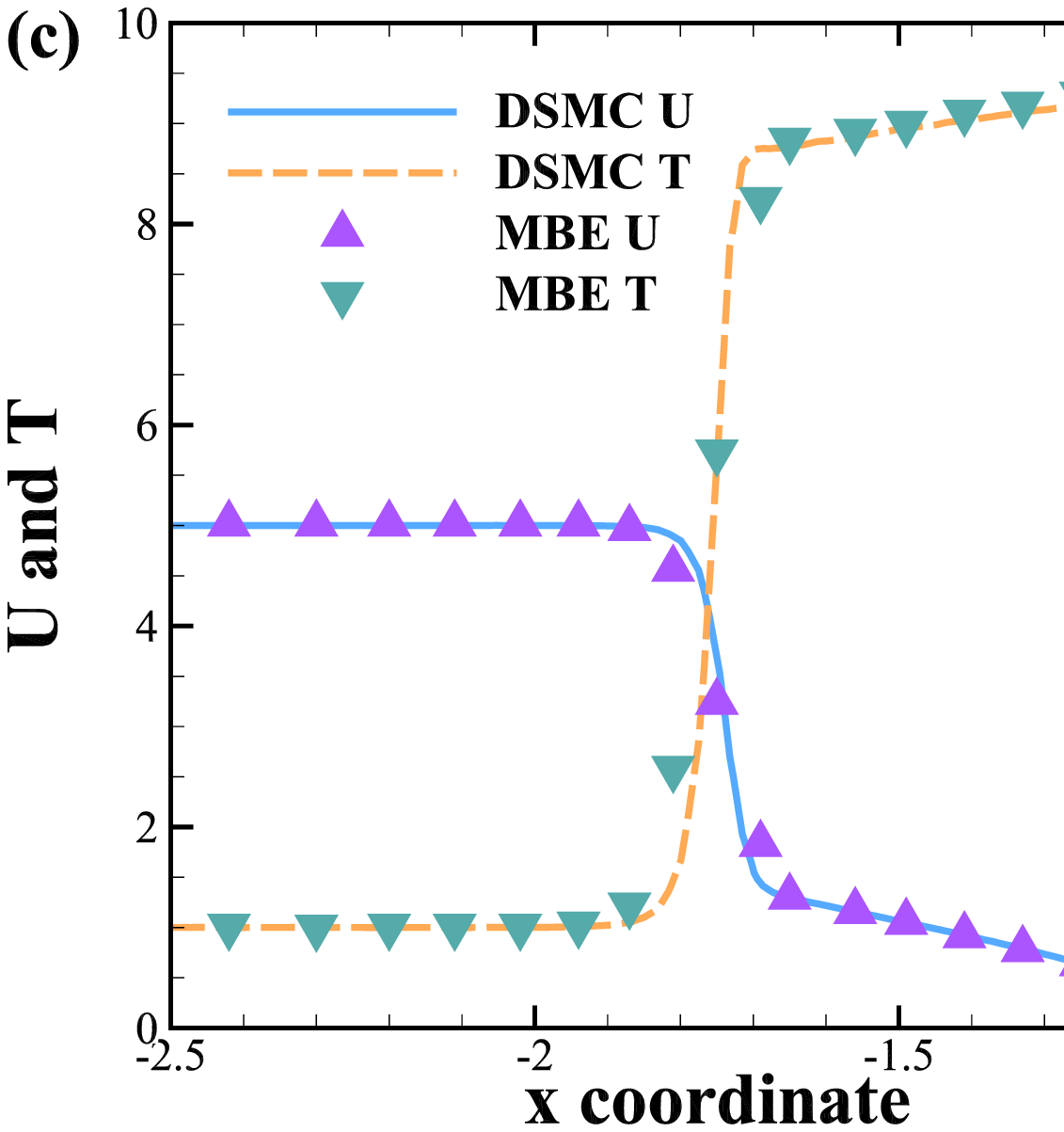}
}\hspace{0.001\textwidth}%
\subfigure{
\includegraphics[width=0.22\textwidth]{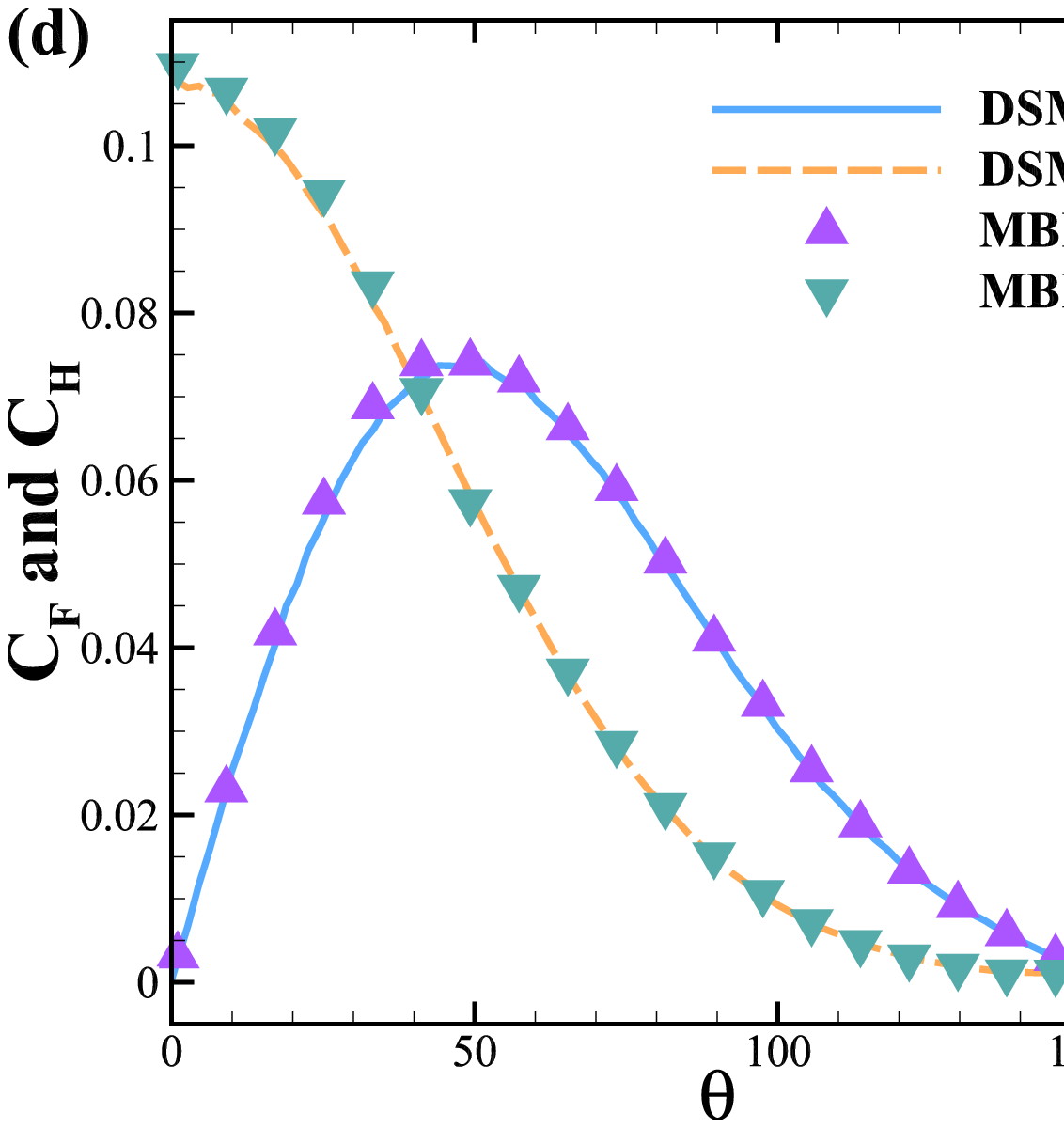}
}
\caption{\label{Fig:fig2} Results of argon gas cylinder flow at $Ma=5$, $Kn=0.01$: (a) Ma contour, (b) macro-weight contour calculated by $\left(1-e^{-\delta t_2/\tau}\right)$, (c) velocity and temperature at the stagnation line, the reference values are $\sqrt{\gamma RT_{\infty}}$ and $T_{\infty}$, respectively, (d) shear stress and heat flux on the solid wall, the reference values are $0.5\rho_{\infty}U_{\infty}^2$ and $0.5\rho_{\infty}U_{\infty}^3$, respectively.}
\end{figure}

A more scale-sensitive case is the hypersonic cylinder flow in a transitional inflow condition\cite{lofthouse,fei} ($Ma = 5$, $Kn = \frac{\lambda}{L} = 0.01$ ($\frac{\lambda}{L} \propto \frac{\tau}{\Delta t}$), the inflow temperature $T_{\infty}$ is the same as the wall temperature $T_w$. The working gas is the argon gas modeled by variable hard sphere potential with $\omega = 0.81$. This setting of working gas is also used in the later test cases. As shown in Fig.\ref{Fig:fig2}, the local physical scale changes significantly in a single flow field. The DSMC solution with extremely small mesh cells (to satisfy the condition $L_{cell}\leq {{\lambda}}/3$) with large computational cost is used as the benchmark solution. The detailed macroscopic variables at the stagnation (central horizontal) line and the sensitive viscous heat flux and shear stress along the cylinder surface are illustrated in Fig.\ref{Fig:fig2}. The results given by the MBE solver match well with the benchmark solution from brute-force computation. Notice that, since the resolution of MBE solver is not limited by the microscopic spatial scale (such as $\lambda/3$), its mesh can be set according to the slops of the flow field. Therefore, the number of collision cells of DSMC is about $2.7$ million, and the cell number of MBE solver is only about $17.9$ thousand in this case. The computational cost of both methods is presented in Tab.\ref{Tab:tab1}.

\begin{table}[htb]
\centering
\caption{Computational cost of the DSMC solver and MBE solver}\label{Tab:tab1}
\begin{tabular}{*{3}{|c|c|c}}
\hline
                                     &DSMC          &MBE        \\   \hline
    Model molecule number (million)  &$21.3$        &$3.3$      \\   \hline
    Mesh number (thousand)           &$2700$        &$17.9$     \\   \hline
    Computation time (hour)          &$122.7$       &$24.7$     \\
\hline
\end{tabular}
\end{table}

\begin{figure}[htb]
\centering
\subfigure{
\includegraphics[height=0.17\textwidth]{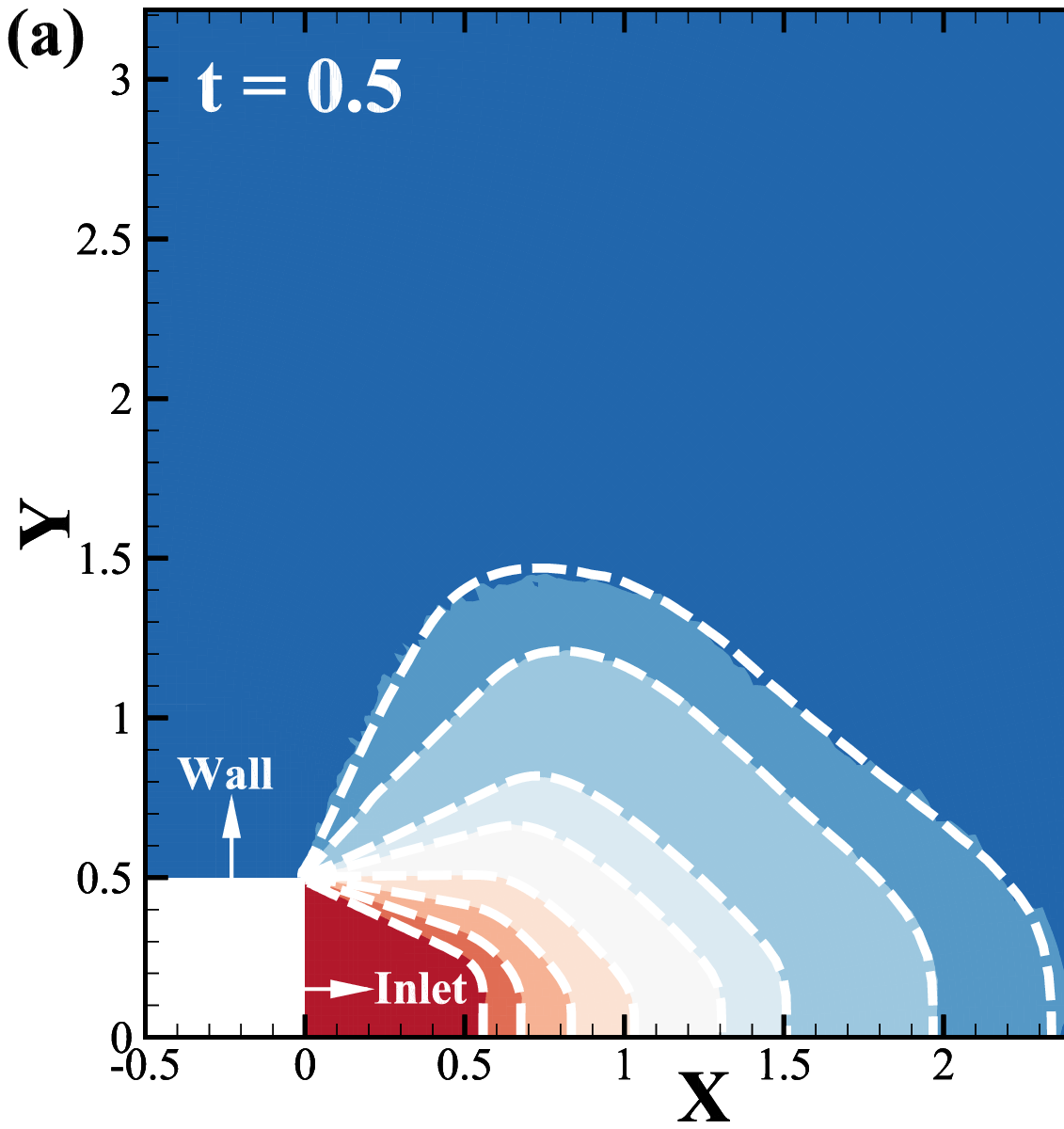}
}
\subfigure{
\includegraphics[height=0.17\textwidth]{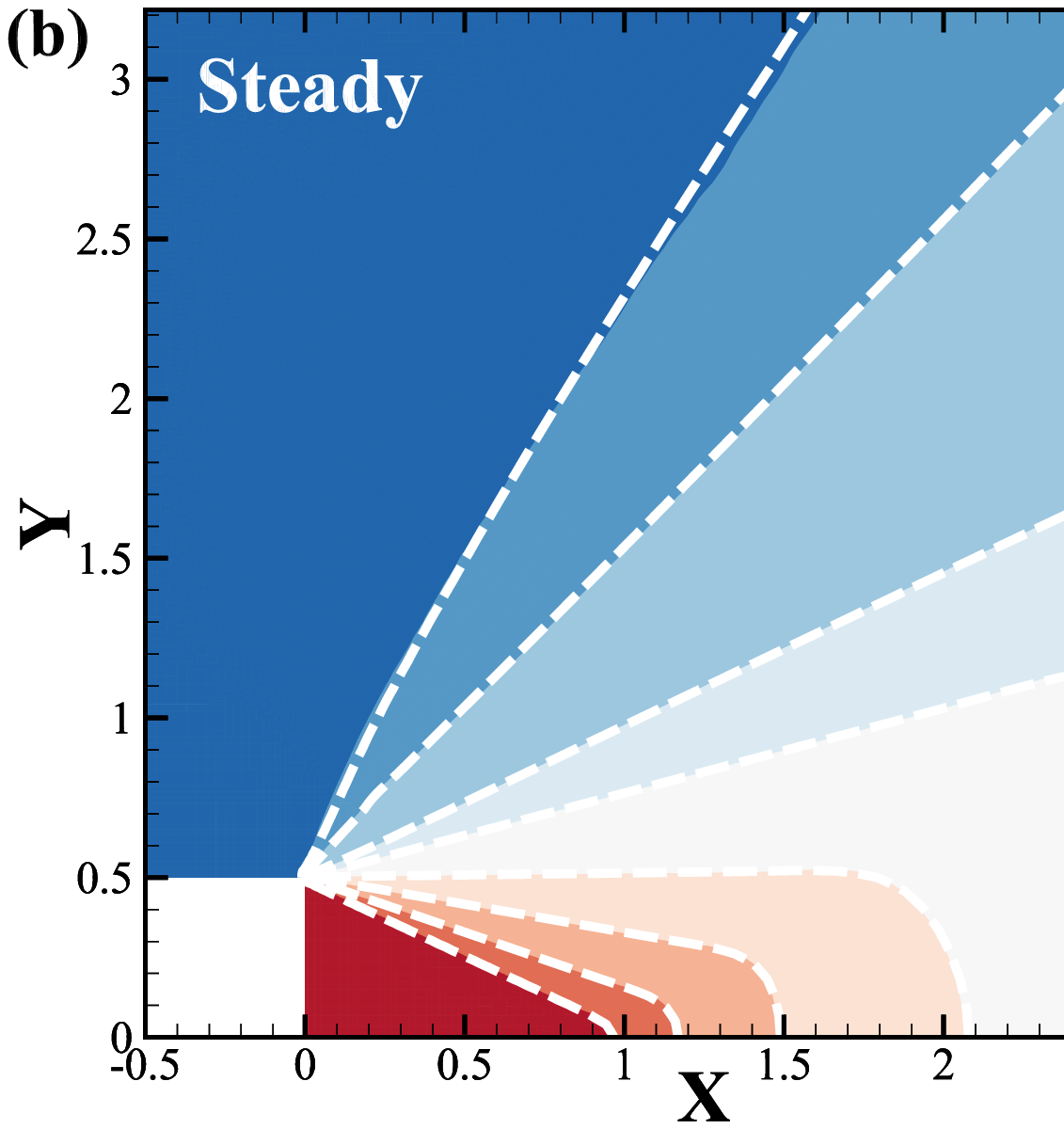}
}
\subfigure{
\includegraphics[height=0.17\textwidth]{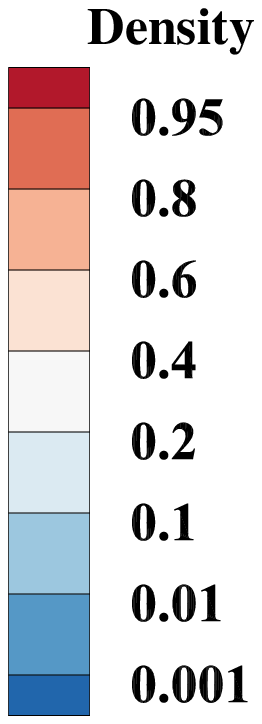}
}
\caption{\label{Fig:fig3} Result of jet flow into vacuum (The density fields predicted by MBE solver are exhibited by contour and benchmark solutions are exhibited by line): (a) Near the start time ($t=0.5$, $t_{ref}=L/\sqrt{RT_{\infty}}$), (b) steady time.}
\end{figure}

A more challenging case is the jet flow into a vacuum environment, which is significant for aerocrafts and micro-electromechanical systems. For this unsteady case, the physical scale not only varies in the flow field but also varies with time. In Ref.\cite{cicp}, this type of flow is specially studied, and a classical condition is applied to validate the present MBE. The inflow Ma is $2.19$, and Kn is $10^{-4}$, whose reference length is the nozzle width. The inflow temperature $T_{\infty}$ is the same as the wall temperature $T_w$). The results near the starting time ($t=0.5, t_{ref}=L/\sqrt{RT_{\infty}}$) and at steady state are shown in Fig.\ref{Fig:fig3}. The density fields predicted by MBE solver (exhibited by contour) are in line well with benchmark solutions (exhibited by dash line).

\begin{figure}[htb]
\centering
\includegraphics[width=0.45\textwidth]{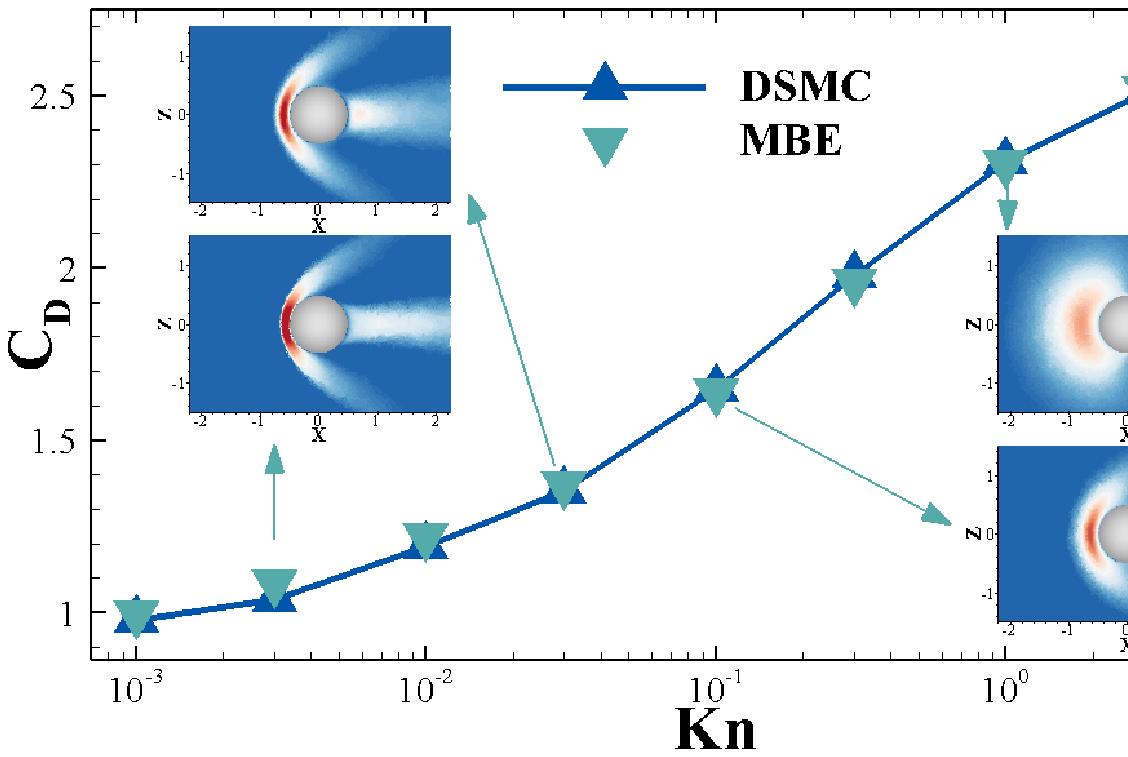}
\caption{\label{Fig:fig4} Drag coefficients of supersonic sphere flow at $Ma=3$.}
\end{figure}

For 3D cases, the hypersonic sphere in all flow regimes is considered. In this case, the inflow Ma is $3$, and Kn spans from $10^{-3}$ to $10$. The inflow temperature is identical to the wall temperature. The drag coefficients are summarized and closely match the results obtained from DSMC simulations, as shown in Fig.\ref{Fig:fig4}. The detailed values are exhibited in Tab.\ref{Tab:tab2}. To obtain a compromised benchmark data, the DSMC is also used in the low Kn case, despite of the huge computational cost.

\begin{table}[htb]
\centering
\caption{Comparison of drag coefficients between the DSMC solver and MBE solver}\label{Tab:tab2}
\begin{tabular}{*{10}{|c|c|c|c|c|c|c|c|c|c}}
\hline
    Kn   &$0.001$ &$0.003$ &$0.01$  &$0.03$  &$0.1$   &$0.3$   &$1$     &$3$     &$10$  \\   \hline
	DSMC &$0.98$  &$1.04$  &$1.19$  &$1.35$  &$1.65$  &$1.98$  &$2.31$  &$2.51$  &$2.63$\\  \hline
    MBE  &$1.00$  &$1.09$  &$1.22$  &$1.37$  &$1.64$  &$1.96$  &$2.30$  &$2.52$  &$2.63$\\  \hline
\hline
\end{tabular}
\end{table}

In summary, a MBE is found for the complex system of neutral gas flows across all flow regimes, which is physically a scale-dependent convex combination of the single-scale BE for rarefied flows and the LBE for near-continuum flows, connecting the aerodynamics with rarefied gas dynamics in a multiscale framework. Benefitting from the clear physical picture of MBE, a concise numerical solver is developed which can be viewed as a unification of the rarefied DSMC solver and the continuum NS solver. The validity of MBE and efficiency of its numerical solver are proved in the above benchmark and challenging test cases. Finally, we hope that the process of finding multiscale master equations for complex physical systems presented in this work will be useful for other areas of multiscale physics and complex science.

This work is supported by National Natural Science Foundation of China (12172301, 12072283).




\FloatBarrier
\nocite{*}
\bibliography{MBEref}

\end{document}